\documentclass[a4paper,11pt]{article}
\pdfoutput=1 

\usepackage{jinstpub} 

\usepackage{lineno}

\usepackage{siunitx}
\DeclareSIPrefix\micro{\text{\textmu}}{-3} 
\usepackage{placeins}
\usepackage[version=4]{mhchem} 
\usepackage{booktabs}

\newcommand{\scarf}{\textsc{Scarf}}
\newcommand{\llama}{\textsc{Llama}}

\newcommand{\lapis}{\textsc{Lapis}}

\title{\boldmath A liquid-phase loop-mode argon purification system}

\author[a, 1]{Christoph Vogl,\note{Corresponding author.}}
\author[a]{Mario Schwarz,}
\author[a]{Patrick Krause,}
\author[b]{Grzegorz Zuzel,}
\author[a]{and Stefan Schönert}

\affiliation[a]{Technical University of Munich, TUM School of Natural Sciences, Department of Physics,\\James-Franck-Str. 1, 85748 Garching, Germany}
\affiliation[b]{Jagiellonian University, M. Smoluchowski Institute of Physics,\\ul. prof. Stanislawa Lojasiewicza 11, 30-348, Cracow, Poland}

\emailAdd{christoph.vogl@tum.de}

\abstract{Noble gas and liquid detectors rely on high chemical purity for successful operation. While gaseous purification has emerged as a reliable method of producing high-purity noble fluids, the requirement for large mass flows drives the development of liquid-phase purification. We constructed a medium-scale liquid argon (LAr) purification system based on a copper catalyst and \qty{4}{\angstrom} molecular sieve capable of purifying \qty{1}{\tonne} of commercial LAr 5.0 to a long effective triplet lifetime of $\tau_3 \sim \qty{1.3}{\micro \s}$. We further demonstrate that a quenched effective triplet lifetime of $\tau_3 \sim \qty{1}{\micro \s}$, due to contamination by air, can be recovered in loop-mode purification to $\tau_3 \sim \qty{1.3}{\micro \s}$ after > 20 volume exchanges.}


\proceeding{LIDINE 2023: LIght Detection In Noble Elements\\
  September 20-22\\
  Madrid, Spain}

\begin{document}
\maketitle
\flushbottom

\section{Introduction}

Impurities deteriorate the performance of noble gas and liquid detectors. Nitrogen and oxygen decrease the light yield of scintillation detectors through quenching processes \cite{acciarriEffectsNitrogenContamination2010, acciarriOxygenContaminationLiquid2010}, oxygen and water absorb VUV photons \cite{acciarriOxygenContaminationLiquid2010, onakaVacuumUVAbsorption1968}, and electronegative impurities capture ionization electrons in time-projection-chambers \cite{bakaleEffectElectricField1976}. The continuous increase in detector size and the resulting need to transport VUV photons and ionization electrons over long distances lead to stringent purity requirements. Purifying noble gases and liquids has become a standard experimental technique in astroparticle physics. 

Most liquid noble gas experiments perform purification in the gas phase with a commercial purifier based on special getter alloys operated at a high temperature of around \qty{400}{\degreeCelsius} \cite{entegrisGateKeeperHGUPurifier2019}. The gas is either drawn from the ullage or evaporated from the liquid phase. It is cleaned in a hot getter, condensed in a heat exchanger, and fed back to the liquid volume. Experiments such as DarkSide-50 \cite{darksidecollaborationLowMassDarkMatter2018} and XENON1T \cite{aprileXENON1TDarkMatter2017} have demonstrated impurity concentrations on the sub-parts-per-billion level in liquid argon and xenon using this technique.

Due to limited throughput and growing detector mass, gaseous purification is reaching its limits, fostering interest in liquid-phase purification where much larger mass flows are possible (the density of, e.g., argon is $\sim780$ times higher in the liquid phase at the boiling point than in the gaseous phase at standard temperature and pressure). Liquid-phase purification is the method of choice for giant liquid argon time-projection-chambers studying neutrinos at reactors and particle beams  \cite{cenniniArgonPurificationLiquid1993, curioniRegenerableFilterLiquid2009, andersonArgoNeuTDetectorNuMI2012, antonelloExperimentalObservationExtremely2014, acciarriLongBaselineNeutrinoFacility2016,  acciarriDesignConstructionMicroBooNE2017}. However, also low-background experiments as XENONnT \cite{planteLiquidphasePurificationMultitonne2022} and \textsc{Legend-200} \cite{haranczykPurificationLargeVolume2022, voglLiquidArgonPurification2021,legendcollaborationLEGEND1000PreconceptualDesign2021} employ liquid-phase purification.

This paper presents the LAr purification system \lapis\ (Liquid Argon Purification Instrument for \scarf\footnote{\scarf\ is the Subterranean Cryogenic ARgon facility \cite{wiesingerTUMLiquidArgon2014}, a \qty{1}{\tonne} LAr cryostat used for R\&D for \textsc{Legend} and previously for \textsc{Gerda}.}) at the shallow underground laboratory of the Technical University of Munich. It is a down-scaled sibling of \textsc{Llars}  (the \textsc{Legend} Liquid ARgon purification System), provides the possibility to test and develop purification strategies on a more accessible scale than in the \qty{90}{\tonne} LAr tank of \textsc{Legend-200}, and supplies high-purity LAr for germanium detector tests, scintillation and doping studies.

\section{Experimental setup}
\label{sec:experimental-setup}

Procurement of high-purity liquid argon is challenging. While LAr can, in principle, be acquired in 5.5 and 6.0 grades, LAr 5.0 is often used instead because of its lower costs and higher availability. LAr 5.0 contains up to \qty{2}{\micro\mol\per\mol} oxygen and \qty{3}{\micro\mol\per\mol} water\footnote{Due to the possible confusions between ppm (part-per-million) by amount and ppm by mass, we chose to denote concentrations by \unit{\micro \mol \per \mol}, identical to ppm by amount (or molar quantity).}, quenching  the argon excimers and leading to a reduction of the effective triplet lifetime to $\tau_3 < \qty{400}{\nano \s}$. \lapis\ is designed to purify up to \qty{1}{\tonne} of LAr 5.0 such that the effective triplet lifetime is not quenched anymore. This corresponds to a value of $\tau_3 \sim \qty{1.3}{\micro \s}$, and oxygen contaminations below \qty{0.01}{\micro \mol \per \mol} \citep{acciarriOxygenContaminationLiquid2010}. Purification is possible in batch mode, i.e., during the filling of \scarf, or in loop mode using a submerged cryogenic LAr pump at the bottom of the cryostat. 

\lapis\ contains two traps. One filled with \qty{2.3}{\kg} Q-5 copper catalyst ($14 \times 28$ mesh) by Research Catalysts \cite{ResearchCatalystInc} to remove oxygen, and the other filled with \qty{2.0}{\kg} \qty{4}{\angstrom} molecular sieve (\qty{1.6}{\mm} to \qty{2.6}{\mm} large beads) from Sigma Aldrich \cite{SigmaAldrich} to remove water. Copper captures oxygen via the reaction \ce{O2 + 2Cu -> 2CuO}, whereas water attaches to the molecular sieve through van der Waals forces. There is limited evidence, currently internal to the \textsc{Legend} collaboration, that these substances also trap nitrogen, albeit with low efficiency\footnote{During the filling of the \textsc{Legend-200} cryostat, one truck delivery contained contaminated LAr with a low triplet lifetime. In an independent test after the filling has been stopped, the \textsc{Legend} LAr purification system was able to significantly increase the triplet lifetime, albeit not satisfactory, proving that the contamination can be at least partly removed. There is evidence that the contamination only entailed nitrogen, although not at high sensitivity. Some of these collaborative efforts have been documented in \cite{voglLiquidArgonPurification2021}. A more detailed analysis is currently being prepared for publication.}.

\begin{figure}
  \centering
  \includegraphics[height=.9\textheight]{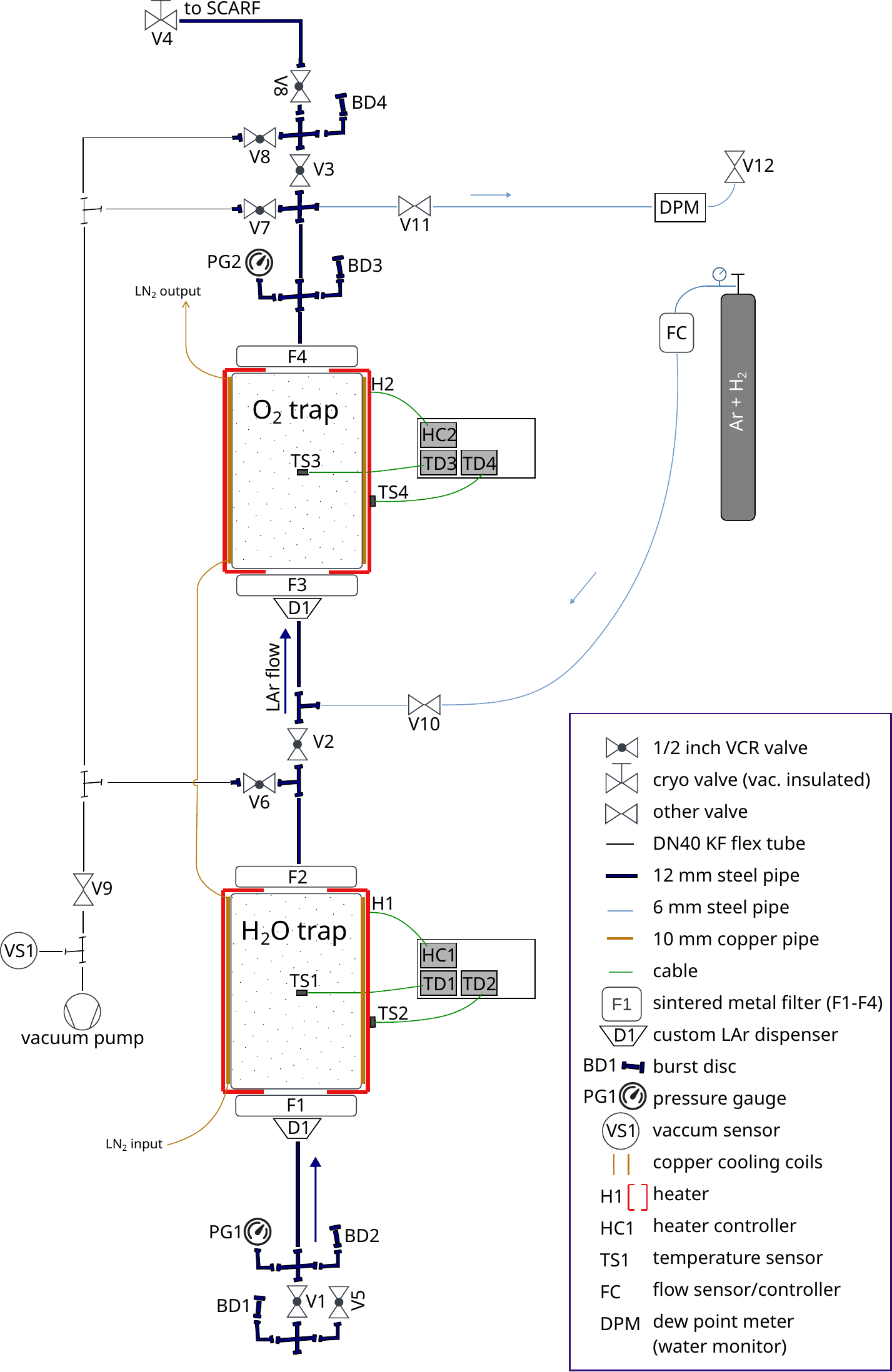}
  \caption{Piping and instrumentation diagram of \lapis, the Liquid Argon Purification Instrument for \scarf. LAr is pushed into the system from the bottom, enters first the water trap, then the oxygen trap, and exits at the top. The equipment for regeneration, i.e., electrical heaters, reduction gas, and a vacuum pump, are also shown. For details, see text.}
  \label{fig:piping-and-instrumentation-diagram}
\end{figure}

The traps are \qty{350}{\mm} long stainless steel cylinders, metal-sealed by CF100 flanges. Sintered metal filters with \qty{10}{\micro \m} pore size set inside the flanges hold the purification substances in place, and custom-made dispensers distribute the LAr flow over the cross-section. \autoref{fig:piping-and-instrumentation-diagram} shows the piping and instrumentation diagram of the system.
The LAr input is located at the bottom valve (V1) and provided either by the LAr pump (loop mode) or by a LAr tank delivery (batch mode). The water trap is located upstream because the oxygen trap is more efficient in the absence of moisture \cite{cenniniArgonPurificationLiquid1993}. The LAr flows upwards (i.e., against gravity) through the purification columns to facilitate a homogeneous distribution over the cross-section of the traps. 

The traps are equipped with electrical heaters and two temperature sensors: One in the center and one on the heater. Before the first use, and when the purifier is saturated with impurities, it must be regenerated. The molecular sieve is cleaned by elevating its temperature to \qty{250}{\degreeCelsius} and applying vacuum. The oxidized copper is heated from \qtyrange{175}{225}{\degreeCelsius} and flushed with hydrogen diluted in an inert carrier gas. Flushing with hydrogen reduces the copper-oxide via $\ce{2CuO + H2 + heat -> 2Cu + 2H2O}$ to pure copper and water. Monitoring the water content in the exhaust permits to determine when the regeneration is complete. The vendor recommends a concentration of hydrogen of \qtyrange{1.5}{2}{\percent} \cite{daveartripReductionGuidelinesCopper2021} to prevent thermal damage during the exothermal reduction reaction, however often concentrations up to \qty{5}{\percent} are used. Initially, we used \qty{5}{\percent} hydrogen in nitrogen but switched later to \qty{2}{\percent} hydrogen in argon. Around 800 trap volumes must be exchanged to guarantee a complete reduction of the copper catalyst using a \qty{5}{\percent} hydrogen mixture \cite{haranczykPurificationLargeVolume2022}. To reduce heat load from the environment, \lapis\ is insulated with rubber foam along the LAr path.

\section{Batch-mode purification}

\lapis\ was used to fill \scarf\ with \qty{1}{\tonne} of LAr from a commercial vendor. We purified two \qty{600}{\liter} tanks containing LAr 5.0 in batch mode in \qtyrange{3}{4}{\hour} each. The initial impurity concentrations were certified by the vendor to be \qty{0.5}{\micro \mol \per \mol} (\qty{0.3}{\micro \mol \per \mol}) oxygen and \qty{0.5}{\micro \mol \per \mol} (\qty{0.2}{\micro \mol \per \mol}) nitrogen for the first (second) tank. According to literature \cite{acciarriOxygenContaminationLiquid2010, acciarriEffectsNitrogenContamination2010}, an effective triplet lifetime of $\tau_3 < \qty{1.0}{\micro \s}$ is expected.
We measure $\tau_3$ of the purified LAr in \scarf\ with the \textsc{Legend} Liquid Argon Monitoring Apparatus, \llama\ \cite{schwarzLiquidArgonInstrumentation2021, schwarzPhDThesisPreparation}, an in-situ device based on a silicon-photomultiplier array developed for and integrated in \textsc{Legend-200} to monitor the LAr optical parameters. \llama\ monitors the LAr scintillating light emitted by an encapsulated $^{241}$Am source. An identical system is available for \scarf. After purification we find $\tau_3 = \qty{1.31(2)}{\micro \s}$, demonstrating the purification capabilities of \lapis.

\section{Loop-mode purification}
After several months of \scarf\ operation, an air leak decreased $\tau_3$ to \qty{0.98(2)}{\micro \s}, and we decided to use \lapis\ to restore the scintillation properties using a cryogenic submersible pump installed at the bottom of the cryostat. It is a custom-built piston pump by the Institute of Air Handling and Refrigeration in Dresden, Germany \cite{ILKDresdenInstitut}, and based on a linear drive. The volume flow can be regulated between \qty{75}{\liter \per \hour} and \qty{495}{\liter \per \hour} by defining the frequency of the piston's movement. To reduce the amount of boil-off gas, especially during the cooldown of the purification system, we installed heat-exchanging copper coils at the traps and cool them down with liquid nitrogen during and before operation.

LAr is extracted from the bottom of \scarf, pumped through \lapis, and reinserted above the liquid. \llama\ is continuously monitoring the optical parameters. Four loop-mode purification runs were necessary to restore $\tau_3$ to its former value of $\sim\qty{1.3}{\micro \s}$. In total, more than \num{20} volumes were exchanged, vastly more than initially expected. Assuming perfect mixing and impurity trapping, a reduction in impurity concentration by $1/\mathrm{e}$ per volume exchange would be expected. 

Assigning the culprit for the inefficient purification is challenging. The contamination is known to be air, which consists almost exclusively of nitrogen, oxygen, argon, and water. Water and oxygen can be efficiently removed from a LAr stream using the materials in \lapis, whereas efficient nitrogen removal has yet to be convincingly demonstrated. It stands to reason that water and oxygen got removed fast, whereas nitrogen was removed slowly. In principle, this could be tested by modeling the time evolution of the scintillation parameters, where a fast initial recovery is expected to be followed by a slow full recovery if \lapis\ captures nitrogen with low but nonzero efficiency. This is, however, outside the scope of this work.

A different interpretation is also possible: Water introduced by the contamination could freeze out onto cold surfaces (e.g., cryostat walls or LN$_2$ cooling coils) and get released slowly into the LAr bulk by diffusion, driven by establishing chemical equilibrium between the solid and liquid phases. 

\begin{figure}
\centering
\includegraphics[height=\textheight/3*\real{1.0175}*\real{0.996}]{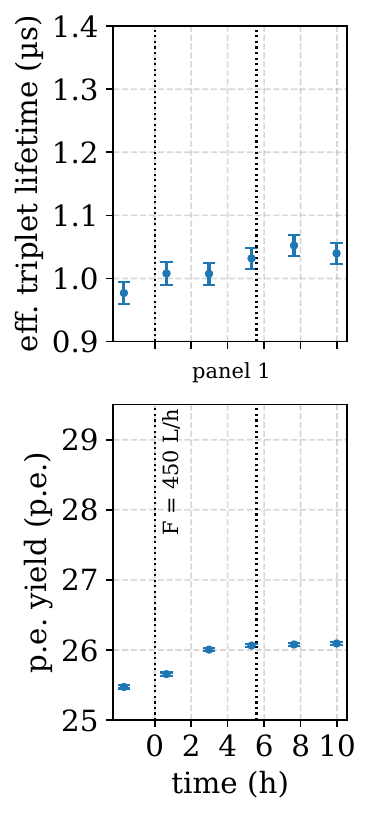}\includegraphics[height=\textheight/3]{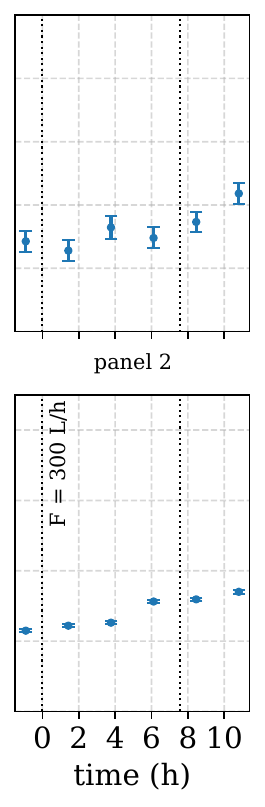}
\includegraphics[height=\textheight/3]{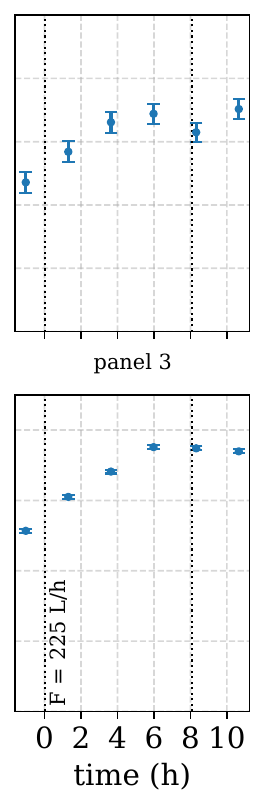}
\includegraphics[height=\textheight/3]{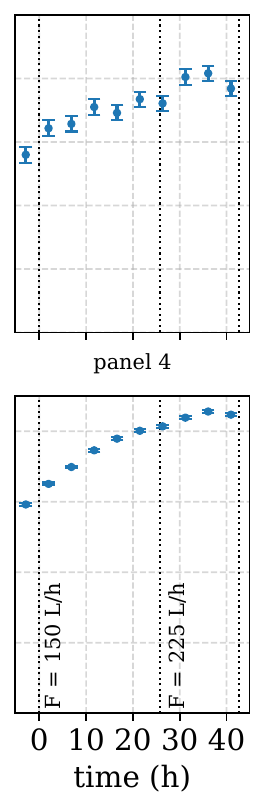}
\includegraphics[height=\textheight/3]{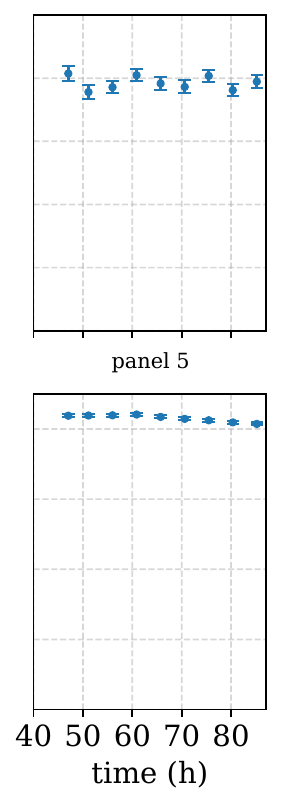}
\caption{Evolution of the effective triplet lifetime (top) and p.e.-yield (bottom) measured by \llama\ during loop-mode purification. The LAr pump runs in the timeframes between two dotted vertical lines and is off in the last panel. The second dotted line in panel four denotes an adjustment of volume flow. The x-axis labels show the time since the pump is running. The purification system was regenerated once at the beginning and a second time between panels one and two. The molecular sieve was regenerated a third time between panels three and four. For details, see text.}
\label{fig:purity-evolution}
\end{figure}

Both hypotheses are supported by limited evidence: \textsc{Legend}-internal observations seem to indicate that nitrogen can be removed from LAr with the present substances, albeit inefficiently, as mentioned in \autoref{sec:experimental-setup}. The frozen water hypothesis receives support in the following paragraphs and could explain the drop in scintillation parameters between panel three and four in \autoref{fig:purity-evolution} as well as the slight decline in p.e.-yield in the last panel. Thus, we are not able to distinguish between these two hypotheses and cannot assign the reason for the inefficient purification with certainty. A combination of both might also be possible.

The first three panels in \autoref{fig:purity-evolution} show the effective triplet lifetime and the photo-electron yield as measured by \llama\ in single-day purification runs where we pre-cooled \lapis\ in the morning, started the pump, purified six to eight hours and turned the system off overnight. Several different volume flows were tested: \qty{450}{\liter \per \hour} in the first panel, \qty{300}{\liter \per \hour} in the second, and \qty{225}{\liter \per \hour} in the third. While the enhancement in scintillation parameters is most pronounced in the third panel with the lowest volume flow, the nonlinear dependence on the impurity concentrations necessitates more careful modeling to determine the optimal purification speed. Before and after the first run, \lapis\ was fully regenerated. The molecular sieve specifically was regenerated another time after the third run.

Panel four shows a \qty{42}{\hour}-long run where we kept the system on continuously. The sudden drop of both scintillation parameters between the end of panel three and the beginning of panel four is currently not understood but might be caused by an inflow of additional impurities at system startup. An alternative explanation would be the abovementioned hypothesis of frozen water present at cold surfaces being released into the liquid. The volume flow is \qty{150}{\liter \per \hour} for the first \qty{28.5}{\hour} and \qty{225}{\liter \per \hour} for the rest. The effective triplet lifetime recovers its original value of $\sim \qty{1.3}{\micro \s}$, whereas the p.e.-yield is only \qty{29.21(2)}{pe} at the end, compared to \qty{34.56(1)}{pe} directly after filling. The origin of this difference is not clear. Quenching can be excluded due to the long effective triplet lifetime, and attenuation in the LAr bulk is unlikely given that the distance between the light source and the light detectors used to extract the p.e.-yield is only a centimeter in \llama. A thin, strongly attenuating layer of frozen water on the surface of the detectors could explain this effect as well.

The last panel of \autoref{fig:purity-evolution} shows the evolution of the scintillation parameters after we shut down the pump. There is no evidence of $\tau_3$ degradation, but the more sensitive p.e.-yield is decreasing slightly, possibly due to residual outgassing or frozen water dissolving into the liquid.

Due to the currently poor thermal insulation of \lapis, LAr was evaporating during the runtime of the pump and released into the atmosphere. We estimate to have lost \qty{365}{\liter} of LAr. This will be mitigated in the future by moving the system into a vacuum-insulated cold box.

\FloatBarrier
\section{Conclusion}
We presented \lapis, the Liquid Argon Purification Instrument for \scarf, and provided two demonstrations of its usage: Batch-mode purification while filling and loop-mode purification thereafter. Liquid argon contaminated with air, featuring a quenched effective triplet lifetime of $\tau_3 = \qty{0.98(2)}{\micro \s}$ was restored with loop-mode purification using dispersed copper and molecular sieve \qty{4}{\angstrom}. $\tau_3$ was raised back to its original value of \qty{1.3}{\micro \s}. However, > 20 volume exchanges were necessary to achieve that, and the p.e.-yield was not fully restored. The experience gathered here provides valuable insight into applying LAr loop-mode purification for the \textsc{Legend} collaboration, which owns a similar but larger system.

\FloatBarrier
\section*{Acknowledgments}
This work has been supported in part by the German Federal Ministry for Education and Research (BMBF) within the project 05A20WO1 and the Polish National Science Centre (Grant No.\ UMO-2020/37/B/ST2/03905).

\bibliographystyle{JHEP}
\bibliography{LIDINE_2023.bib}

\end{document}